\documentclass[usenatbib]{mn2e}
\usepackage{longtable}
\usepackage{graphicx}
\usepackage{amsmath}
\usepackage{amssymb}
\usepackage{times}
\usepackage{mathtools}

\usepackage{morefloats}
\usepackage{etex}

\pdfoutput=1

\title[Radial distributions of multiple populations]{G2C2 - IV: A novel approach to study the radial distributions of multiple populations in Galactic globular clusters }  
\author[J. Vanderbeke et al.]
{Joachim Vanderbeke$^{1}$\thanks{E-mail: Joachimvanderbeke@gmail.com}, Roberto De Propris$^2$, Sven De Rijcke$^{1}$, Maarten Baes$^{1}$, \newauthor Michael West$^{3}$, Javier Alonso-Garc\'ia$^{4,5}$, Andrea Kunder$^{6}$ \\ \\ 
$^{1}$ Sterrenkundig Observatorium, Universiteit Gent, Krijgslaan 281 S9, B-9000 Gent, Belgium\\
$^{2}$ Finnish Centre for Astronomy with ESO (FINCA), University of Turku, V{\"a}is{\"a}l{\"a}ntie 20, FI-21500 Piikki{\"o}, Finland \\
   $^{3}$ Maria Mitchell Observatory, 4 Vestal Street, Nantucket, MA 02554, USA \\
$^{4}$  Instituto de Astrof\'{i}sica, Facultad de F\'isica, Pontificia Universidad Cat\'{o}lica de Chile, Av. Vicu\~{n}a Mackenna 4860, 782-0436 Macul, Santiago, Chile \\
   $^{5}$ The Milky Way Millennium Nucleus, Av. Vicu\~na Mackenna 4860, 782-0436 Macul, Santiago, Chile\\
$^{6}$   Leibniz-Institut f{\"u}r Astrophysics (AIP), An der Sternwarte 16, 14482 Potsdam Germany \\
}

\begin{document}

\date{Accepted. Received }

\maketitle

\label{firstpage}

\begin{abstract}
We use the HB morphology of 48 Galactic GCs to study the radial distributions of the different stellar populations
known to exist in globular clusters. Assuming that the (extremely) blue HB stars correspond to stars enriched in Helium and
light elements, we compare the radial distributions of stars selected according to colour on the HB to trace the distribution
of the secondary stellar populations in globular clusters. Unlike other cases, our data show that the populations are well
mixed in 80\% of the cases studied. This provides some constraints on the mechanisms proposed to pollute the interstellar
medium in young globular clusters.
\end{abstract}


\section{Introduction}\label{sec:introduction}

Globular clusters are now known to host multiple stellar populations, differing in their light element abundances, although iron and heavier elements tend to be largely homogeneous (e.g., \citealt{Gratton2012} for a review). 
It is likely that production of light elements is accompanied by enrichment in Helium as well, where Na-rich stars from the second generations
are more He-rich than the Na-poor and O-rich first generation (FG) stars \citep[e.g.][]{Carretta2007}. The secondary stellar generations (SG) must have 
been formed during a later starburst, from material polluted by the ejecta of massive stars in the original stellar population. Candidate polluters 
include: intermediate mass AGB stars \citep{Izzard2006,Dantona2007}, fast rotating massive stars \citep{Decressin2007,Krause2012,
Krause2013}, ejecta from massive binaries \citep{DeMink2009} and early disk accretion \citep{Bastian2013} where the ejecta from massive 
stars are accreted onto forming low-mass protostars at early times.

It is generally difficult to distinguish between these scenarios purely on the basis of the observed abundance patterns, as they all 
share common characteristics, such as pollution by the products of hot bottom burning in massive systems. However, it may be 
possible to constrain these models via their effects on the radial distribution of polluted stars, i.e, the first and later generations. In
general, we expect that more enriched stars will tend to reside closer to the cluster centres, as the gas needed to fuel star formation
tends to sink to the cluster core (e.g., \citealt{Dercole2008}). In the early disk accretion model, the protostellar disks accrete material
as they travel through the cluster and therefore one expects a somewhat broader distribution. On the other hand, the He-enhanced
stars may lose more mass during the RGB phase and diffuse outwards because of two-body relaxation, yielding a more extended
distribution \citep{Carretta2009}. The radial distribution of stars in clusters may be altered by dynamical evolution. However, simulations 
by \cite{Vesperini2013} predict that in many Galactic clusters the second generations should still be more centrally concentrated than the 
first generation. Opposite, \cite{Decressin2008,Decressin2010} argue that any original radial difference is erased after a Hubble time 
because the relaxation times are much shorter than the ages of globular clusters.

Observationally, \cite{Lardo2011} have used the $u-g$ colour to separate red giants belonging to each population (cf. \citealt{Milone2008})
and show that the UV-faint, Na-rich second generation stars are more centrally concentrated. \cite{Carretta2009} divide their spectroscopic sample into
primordial, intermediate and extreme stars based on the degree of enrichment in Na and O abundances and show that the intermediate 
stars are more centrally concentrated, followed by the primordial and then the extreme subpopulations, although \cite{Lardo2011} criticise
the {\it ad hoc} selection of targets. In our previous studies we have shown how selecting red giants in the crowded central regions of
globular clusters can be difficult and how their photometry may be doubtful, especially from small telescopes and in mediocre seeing 
conditions \citep{Renzini1998,Vanderbeke2014a,Vanderbeke2014b}. It is perhaps suggestive that the two least crowded systems in \cite{Lardo2011}
are those that show no evidence of radial gradients in the distribution of stellar populations. Other studies argue that second generation
stars may reside closer to cluster centres than first generation stars \citep[e.g.][]{Sollima2007,Carretta2010d,Kravtsov2011,Johnson2012},
still reflecting the initial segregation. However, \cite{Dalessandro2014} combined optical and near-UV photometry to study the sub-giant and 
red giant branches of NGC~6362. They concluded that the FG and SG stars share the same radial distribution, making it the first system 
where stars from different populations are found to be completely spatially mixed.

Previous work has used RGB stars as tracers, either from their colours or direct spectroscopy \citep[e.g.][]{Sbordone2011}. Here we propose a different approach,
based on the distribution of stars on the horizontal branch (HB). The dominant parameter determining the HB morphology is the metal abundance \citep[e.g., ][]{Arp1952,Sandage1953}, while other parameters like age, Helium enrichment and mass loss, to name a few, may produce anomalously red or blue HBs (the ''second parameter problem'' --
e.g., \citealt{Dotter2010}). 
Helium abundance variations are linked to the observed light element enhancements for the multiple
generations in globular clusters \citep[e.g.][]{Carretta2006,Dantona2007, Salaris2008,Villanova2009, Marino2011,Villanova2012, 
Monelli2013,Salaris2013,Milone2014a,Gratton2014a,Mucciarelli2014,Gratton2015,Milone2015} and may be taken to trace abundance variations in CNONa, even if these by themselves may not affect the morphology of the HB. 

\cite{Carretta2010b} show that the maximum temperature of the HB correlates
with the range of Na/O abundance in red giants. \cite{Dantona2005} and \cite{Iannicola2009} compare their data on NGC~2808 with 
synthetic models and demonstrate how the He abundance increases bluewards along the HB in their colour-magnitude diagrams (CMDs). The models 
of \cite{Joo2013} also link the HB morphology directly to He enrichment and the presence of multiple stellar generations, although this is a 
2nd order effect on the dominant metallicity which is the 1st parameter affecting the HB. \cite{Marino2011} find that in M4 the red HB (RHB) stars
are O-rich and Na-poor, while the blue HB (BHB) stars are O-poor and Na-rich; the abundance patterns suggest that the elements were produced during hot bottom burning via the CNO cycle and its high temperature NeNa and MgAl branchings \citep[e.g., ][]{Clayton1968} 
and the resulting stars will also be Helium enhanced. Therefore, the distribution of stars on the blue HB within each cluster 
may be used as a proxy for the distribution of stars belonging to each generation.

We exploit the relation between light element enrichment and HB morphology, mediated by He abundance, to study the spatial distribution of stellar populations selected on the HB.
We essentially take temperature 'cuts' along the HB and argue that these correspond
to increasing contributions from He-enriched secondary generations of stars.  For example, \cite{Iannicola2009} can separate the HB of 
NGC\,2808 into three groups, each assumed to correspond to the three (or more) main sequences observed by \cite{Piotto2007}. Observations 
in near ultraviolet generally show numerous gaps on HBs \citep{Iannicola2009}, even if these are continuous in the optical, and these may be used to separate the 
multiple stellar generations.  \cite{Iannicola2009} selected about 2000 HB stars based on both ground-based and HST observations of NGC~2808.
Relative fractions of cool, hot and extreme HB stars do not change radically along the radial profile of the cluster. Therefore, this is evidence 
against the presence of radial differences between stellar subpopulations with different He abundances. \cite{Kunder2013a} studied the cumulative 
fraction of extreme HB stars in M~22 (NGC~6656) but could not draw strong conclusions regarding differences in the radial distributions of blue 
and extremely blue HB (EHB) stars. \cite{Gratton2014a} studied M~22's HB spectroscopically and find some suggestive evidence that SG stars are more 
concentrated than FG stars. However, their study does not cover the extreme HB stars, due to the restrictions on the temperature range.

Here we use our Galactic Globular Cluster Catalog (G2C2) photometry (\citealt{Vanderbeke2014a,Vanderbeke2014b}, see Vanderbeke et al.
2015, in preparation, for a discussion of the colour-magnitude diagrams of all clusters in the G2C2 sample) to address this issue. Our large 
and homogeneous photometry is sufficiently deep that we can trace the HB to the level of the main sequence turnoff in almost all 
clusters and allow us to study the radial distributions of stars at distances beyond several core radii from the cluster centre in many instances (the CTIO FOV is 13.6\arcmin on the side and has a resolution of 0.396\arcsec per pixel). 
The main goal of this study is
to use the distribution of stars on the HB as a proxy for the distribution of stellar populations and use these stars to trace the radial
density profile of the various stellar generations. We wish to test whether second generation stars are more centrally concentrated
than the more primordial objects and if possible consider the effects of dynamical evolution over the past Hubble time. Our homogeneous 
and wide-field photometry proves ideal for this task. 

\section{Sample selection and methodology}\label{sec:sampleselection}
The basic data for this paper derive from our homogeneous photometry of Galactic globular clusters in Papers I and II \citep{Vanderbeke2014a,
Vanderbeke2014b}. The full colour-magnitude diagrams for these clusters will be fully discussed in a subsequent publication. 

\begin{figure*}
\centering 
\includegraphics[scale=0.87,trim= 2.7cm 13.2cm 9cm 5.8cm] {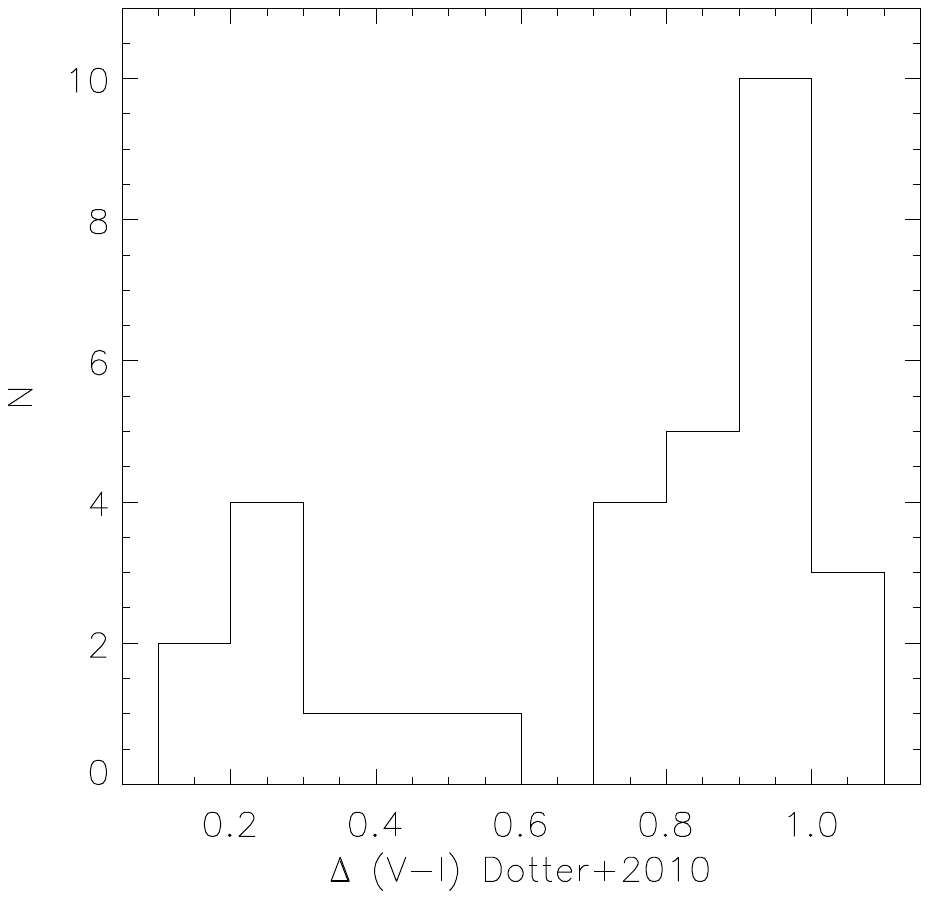}\includegraphics[scale=0.87,trim= 2.7cm 13.2cm 0 5.8cm] {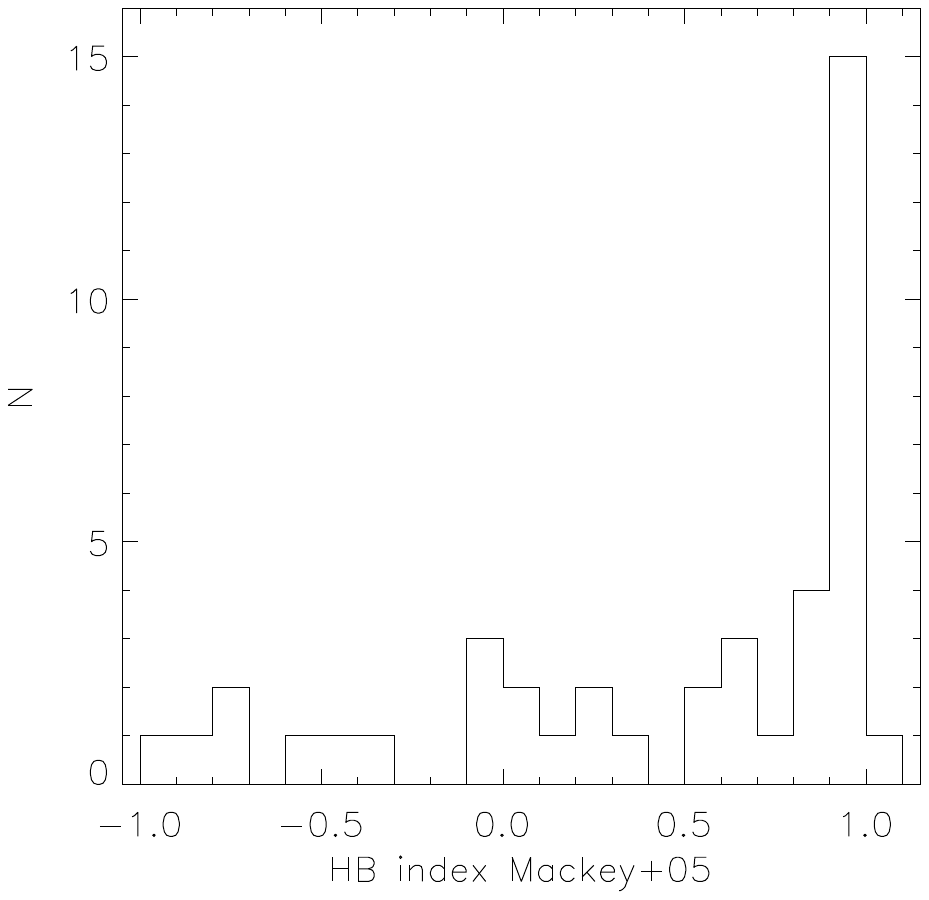} 
\caption{Histograms of the HB indices for our sample from \citealt{Dotter2010} (left) and \citealt{Mackey2005} (right). }
\label{fig:histogram_HB_indices}
\end{figure*}

Because we are interested in identifying objects with extended HBs, and using colour cuts on the HB to select stars in temperature ranges
so that each group contains stars from the primordial or secondary generations, we focus on clusters whose HBs have blueward extensions \citep[see e.g.][]{Mackey2005, Lee2007, Dotter2010}. In Fig.~\ref{fig:histogram_HB_indices} we present histograms of the HB indices of the GCs studied in this
paper. We included all GCs where the HB index is larger than 0.7, whenever good quality data (deep CMDs, low contamination) was available.

While blue stragglers may contaminate counts of extreme HB stars, in most cases only the hottest and more massive blue stragglers would be confused with HB stars, and these are generally very few in number. 
The location of the field stars in the CMD varies from cluster to cluster. However, the coinciding HB region would appear less centrally concentrated by the field star contamination.

Some clusters with rather red HB morphologies were also included, because these GCs either belong to a second parameter pair (e.g. 
NGC~362, NGC~6171), or there is existing evidence of multiple populations with differing radial distributions  
(e.g. NGC~104, NGC~6362), or visual inspection revealed some BHB/EHB stars though the HB is strongly dominated by the RHB (e.g. NGC~1261).

After correcting our CMDs for foreground extinction using the \cite{Cardelli1989} reddening law (using $E(B-V)$ values from the 2010 version of \citealt{Harris1996}), we subdivide the HB in ranges, according to colour, where the first or second generation stars are expected to prevail (cf., \citealt{
Iannicola2009,Marino2011}). This is unfortunately not straightforward when using optical colours. \cite{Joo2013} predict different colours 
for the different generations residing in M~22 (NGC6656) and NGC1851 and show that the generations can have partly overlapping 
colour and magnitude ranges. Therefore, every possible cut will only be an approximate separation of the populations. In this study we 
make the assumption that the majority of RHB/BHB stars represent the stars with primordial abundances, while the majority of EHB 
stars correspond to the enriched population. Either sample is of course contaminated to some extent.

\begin{figure*}
\centering 
\includegraphics[scale=1,trim=8cm 13.5cm 9cm 6cm] {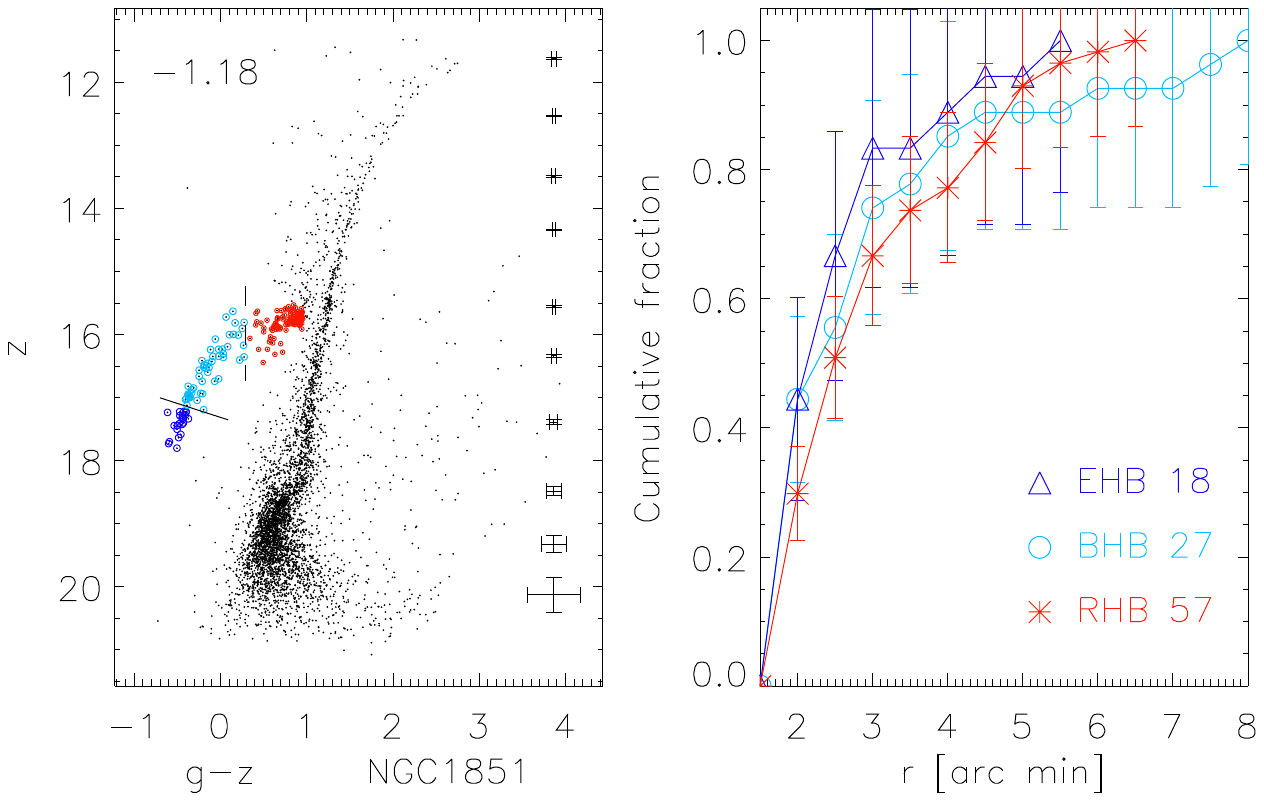}
\caption{NGC1851: Left panel: a $gz$ CMD of NGC 1851 from our data, corrected for foreground extinction (\citealt{Cardelli1989}, using $E(B-V)$ from the 2010 version of \citealt{Harris1996}). The errors shown are the median photometric errors. These errors are the errors given by \textsc{daophot} \citep{Stetson1987} and do not include the photometric calibration errors or the systematic photometric uncertainty found in Paper~I (as these would dominate the error budget). We systematically exclude stars with a magnitude error larger than 1. Right panel: cumulative fractions 
of the radial distributions of the HB regions. The number of stars within the inner and outer limiting completeness radii are given for the different HB regions. }
\label{fig:NGC1851cfpaper}
\end{figure*}

For the sake of homogeneity, we define colour cuts which we will use systematically for all clusters using NGC~1851 to illustrate the different 
steps in our analysis (see Fig. \ref{fig:NGC1851cfpaper}). This cluster is known to host multiple stellar generations and is famous for its unusual bimodal HB morphology, double RGB 
and SGB, possible [Fe/H] spread, CN bimodality and variations in light and s-process element abundances \citep{Saviane1998,Walker1998,
Joo2013}. \cite{Carretta2011} demonstrate that the metal-poor stars in their sample are more concentrated than the metal-rich stars.

\begin{figure}
\centering 
\includegraphics[scale=0.87,trim= 2.7cm 13.2cm 0 5.8cm] {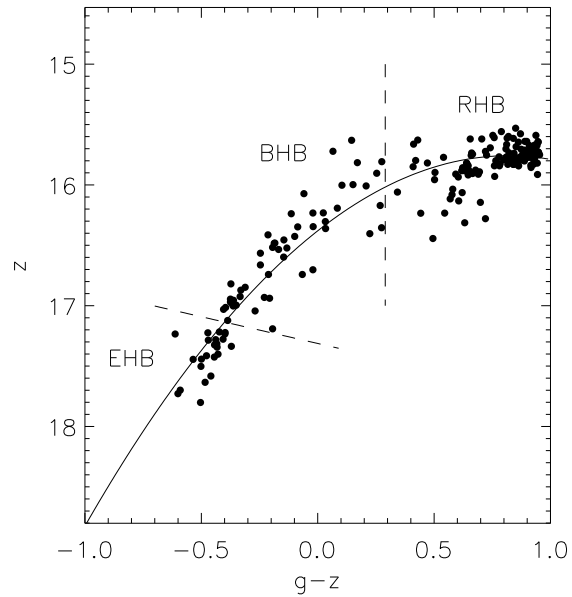}
\caption{Zooming in on the HB of NGC1851. The solid line is a quadratic fit to the HB stars. The dashed lines are the colour cuts applied to separate RHB (cooler/redder than $g-z=0.29$, corresponding to $\sim7000K$), BHB and EHB (hotter/bluer than the dashed line passing through $g-z=-0.4$ (corresponding to $\sim11000K$), which is perpendicular to the solid line) stars. }
\label{fig:NGC1851zoomselection}
\end{figure}

Now we define colour cuts for the different HB subgroups. \cite{Dalessandro2011} showed the temperature distribution of NGC~2808 HB stars and found an effective temperature of about 7000 K as the limit between RHB and BHB stars, close to the blue edge of the RR Lyrae instability 
strip \citep{MoniBidin2012}. This temperature corresponds to a black body colour of $g-z\sim0.29$ and was used as a vertical cut to separate 
RHB and BHB stars. The gap between BHB and EHB stars is located at about 20000K \citep{Dalessandro2011,MoniBidin2012}. However, this 
$T_{eff}$ corresponds to a black body colour of $g-z\sim-0.9$, which is bluer than any of the HB stars in our CMDs. \cite{Dalessandro2011} 
demonstrated that the EHB temperatures derived from a combination of optical filters only can be underestimated by 10000K and more (because
HB sequences with different initial He mass fraction overlap for $T_{eff} > 10 000 K$). Therefore, we are forced to introduce an arbitrary colour cut below the 'bend' of the HB at $g-z\sim-0.4$, corresponding to an effective temperature of 11000K. Because the HB becomes vertical at blue colours, we do not make a pure "vertical" colour cut for the EHB stars. We fit a second degree polynomial to the HB stars and determine the line 
perpendicular to this polynomial passing through $g-z\sim-0.4$. In Fig.~\ref{fig:NGC1851zoomselection} we zoom in on the HB region of 
NGC~1851. We illustrate the fitting procedure by the solid line and the different cuts by the dashed lines. 

\begin{figure*}
\centering 
\includegraphics[scale=1,trim=4.5cm 13.cm 9cm 5.5cm]{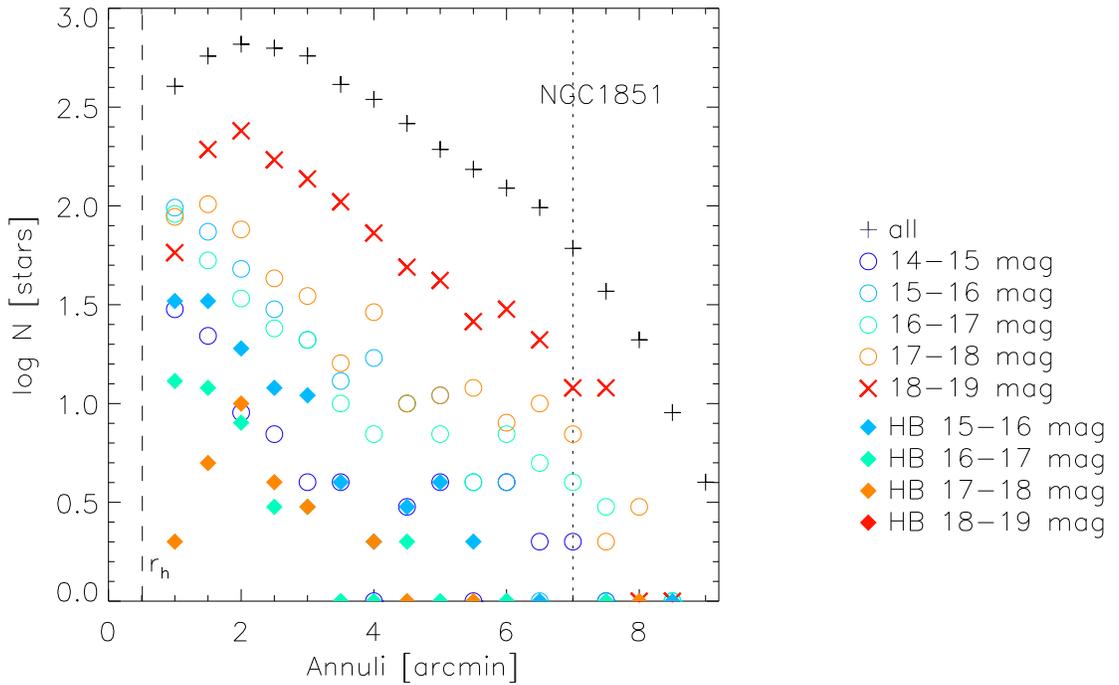}
\caption{NGC1851: radial distribution of the number of stars contained in 0.5\arcmin annuli. The half-light radius $r_h$ is indicated with dashed line, the dotted line represents the radius where part of the annulus is outside the CTIO field of view. }
\label{fig:NGC1851radpaper}
\end{figure*}

Our CMDs are derived from data observed with small ground-based telescopes (both Sloan Digital Sky Survey (SDSS) and Cerro Tololo Inter-American Observatory 
(CTIO) 0.9m telescope), as presented in \cite{Vanderbeke2014a}. Therefore, these suffer from incompleteness, especially close to the cluster centre. We
show the number of stars as a function of distance from the cluster centre in $0.5'$ annuli as a function of luminosity in
Fig.~\ref{fig:NGC1851radpaper}. The HB spans a magnitude range $15.5 < z < 18$, so we need similar completeness levels for 
that magnitude range. Therefore, we choose to consider only the HB stars beyond an inner limiting radius of 1.5\arcmin. In 
Fig.~\ref{fig:NGC1851radpaper}, the half-light radius $r_h$ is indicated with dashed line, the dotted line represents the radius where 
part of the annulus is outside the 13.6\arcmin~CTIO field of view. Note that the inner radial limit is well beyond the half-light radius. 
As the outer limiting radius, we choose 8\arcmin~(hence including almost the entire CTIO field of view). 

In general, we use all stars to determine the completeness at each radius. However, the total number of stars is dominated by the numerous 
RGB stars, while the focus of our study is on the HB stars, which are much bluer. Therefore, we still need to check that the total number of stars is 
representative to make the completeness cuts for the HB, so, that no bias for blue stars was introduced by the characteristics of the CCD. 
The filled symbols in Fig.~\ref{fig:NGC1851radpaper} represent the HB stars within the given magnitude ranges. 
HB stars of different magnitude ranges within the previously determined inner and outer limits show similar completeness levels (although they suffer from low number statistics). Similar conclusions were drawn for another dozen clusters spanning a variety of globular cluster properties (including mass, HB morphology, etc.). 
Therefore, we conclude that we can safely use all the stars, even  if dominated by the RGB stars, to determine 
the completeness as a function of radius. 

We define and colour-code the different HB regions in the left panel of Fig.~\ref{fig:NGC1851cfpaper} and compare the cumulative radial 
distributions of the stars between the inner and outer completeness radii in the right panel of the same figure. Similar figures for all other 
GCs can be found in the online appendix. 

We perform a two-sided KS test to compare the radial distributions of the HB stars as defined above (RHB, BHB and
EHB). The test returns no significant difference between the radial distributions of RHB, BHB and EHB stars. 
In Table~\ref{tab:KS} we show an extract of the KS statistics to guide the reader. 
The complete table can be found in the online appendix. 
Note that our sample also includes $\omega$~Cen (NGC~5139), an object that it is much more complex than a typical mono-metallic globular cluster \citep[e.g.,][]{Villanova2014}. A more detailed analysis of effects of the [Fe/H] spread on the HB selection criteria is beyond the scope of this study. 

\begin{table*}
\centering
\caption{\label{tab:KS} Extract of the KS statistics table for radial distributions of different HB regions. HB$_1$ and HB$_2$ denote the considered HB regions, as defined in the linked figure. N$_1$ and N$_2$ give the number of the stars in both HB regions. D$_{KS}$ gives the Kolmogorov-Smirnov statistic and Prob$_{KS}$ presents the significance level of the KS statistic. Small values show that the cumulative radial distribution of HB$_1$ stars is significantly different from HB$_2$ stars. The last column indicates if the CMD is based on CTIO or SDSS data.}
\begin{tabular}{lcccccccc}
\hline
ID & HB$_1$ & HB$_2$ & N$_1$ & N$_2$ &D$_{KS}$&Prob$_{KS}$&Fig.&\\ \hline

    NGC104 &   red &  blue & 245 & 230 &  0.198 &  0.000          & \ref{fig:NGC104} & CTIO \\
    \hline
    NGC288 &   BHB &   EHB &  17 &  67 &  0.345 &  0.061          &  & CTIO \\ 
    \hline
    NGC362 &   BHB &   EHB &  10 &  26 &  0.223 &  0.814          &  & CTIO \\ 
    NGC362 &   RHB &   EHB &  67 &  26 &  0.372 &  0.008          & & CTIO \\ 
    NGC362 &   RHB &   BHB &  67 &  10 &  0.301 &  0.344          &  & CTIO \\ 
    \hline
   NGC1261 &   BHB &   EHB &   9 &   6 &  0.556 &  0.140         &  & CTIO \\ 
   NGC1261 &   RHB &   EHB &  93 &   6 &  0.226 &  0.897         &  & CTIO \\ 
   NGC1261 &   RHB &   BHB &  93 &   9 &  0.459 &  0.043         &  & CTIO \\ 
    \hline
   NGC1851 &   BHB &   EHB &  27 &  18 &  0.222 &  0.603         & \ref{fig:NGC1851cfpaper} & CTIO \\
   NGC1851 &   RHB &   EHB &  57 &  18 &  0.202 &  0.583         & \ref{fig:NGC1851cfpaper} & CTIO \\
   NGC1851 &   RHB &   BHB &  57 &  27 &  0.183 &  0.526         & \ref{fig:NGC1851cfpaper} & CTIO \\
    \hline
   
\end{tabular}
\end{table*}

\cite{Saviane1998} also studied NGC~1851 and found evidence that the radial distribution of the blue HB stars is different from that of the 
red HB and sub-giant branch stars. Their Fig. 11 shows that blue horizontal branch stars are more centrally concentrated than red horizontal 
branch stars, while our figure suggests that only the EHB stars are slightly more concentrated than the RHB stars. Our KS statistics also indicate 
that the difference in cumulative radial distributions is not very significant. \cite{Saviane1998} also presented HST imagery for the inner 25 arc
seconds. These data did not suggest any significant radial difference for the blue and red HB stars. \cite{Milone2009} also studied NGC~1851 
and did not find any radial stellar population gradients, in agreement with our results. 

\section{Results}

\begin{figure}
\centering 
\includegraphics[scale=0.87,trim= 2.7cm 13.2cm 0 5.8cm] {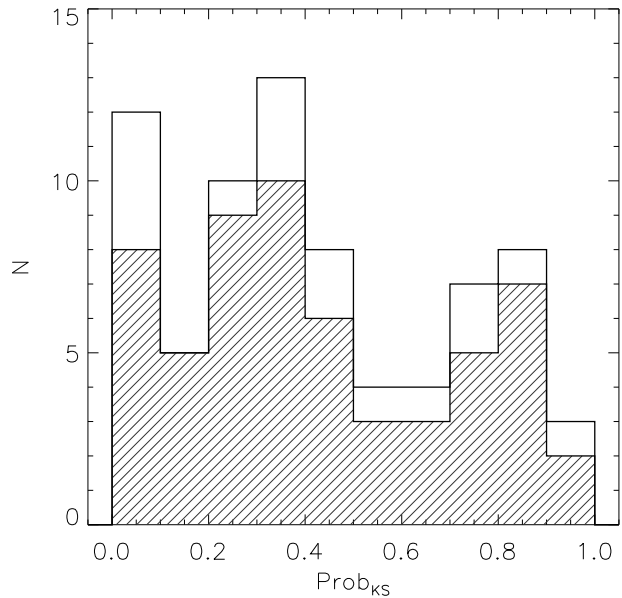} 
\caption{Histogram of the KS probabilities. The histogram without dashes presents the distribution of all clusters, while the dashed histogram represents the subsample for which we consider the EHB. See Table~\ref{tab:KS} for the specific values.}
\label{fig:KS_HB_hist}
\end{figure}

In Fig.~\ref{fig:KS_HB_hist} we show the distribution of the KS probabilities (from CTIO and SDSS data). The distribution of the KS 
probabilities relative to the EHB is given by the dashed histogram. The distribution is trimodal: in some clusters there are strong differences
in the radial distribution of HB stars within each colour range; others show either mild or no differences. 

More than 80\% of the clusters do not show evidence for different radial distributions along their horizontal branch. This suggests that the different stellar populations have similar radial distributions once on the HB, at least for the considered radii, in contrast to the study by \cite{Lardo2011} who probed the red giant branch stars. Moreover, we find no significant radial 
distribution difference for several clusters for which \cite{Lardo2011} found differences in the radial distributions of first and second generation
stars: NGC~5024, NGC~5904, NGC~6205, NGC~6341 and NGC~7089 (M~2). If the position of stars on the HB is
related to their He abundance (and indirectly to the CNONa anomalies), this result is puzzling. In agreement with \cite{Lardo2011} we
find no difference in the radial distributions of stars for NGC~5466, while in NGC~5272 we only find a small difference in radial distributions
(Prob$_{KS}\sim0.2$ for the EHB comparisons), while \cite{Lardo2011} found a significant difference. For NGC~7078 we find a difference 
in the radial distributions for both the RHB-EHB and RHB-BHB comparisons as do \cite{Lardo2011}. For NGC2808 we do not find a radial 
difference between RHB and EHB, in agreement with \cite{Iannicola2009}. For NGC~6362, we found that RHB 
and BHB stars have the same radial distribution, agreeing with \cite{Dalessandro2014}. However, the EHB stars are somehow more 
concentrated than the RGB stars (with Prob$_{KS}\sim0.13$). 

NGC~288, NGC~362 and NGC~6218 are particularly interesting objects (all with $[Fe/H]\sim-1.3$), as these are the first known systems for 
which the second generation appears less concentrated than the first generation, based on our data. Photometric or spectroscopic follow-up 
studies are needed to confirm or disprove these radial distributions.  Disc shocking may be a main contributor to the peculiar radial distributions 
of the different populations in NGC~288 \citep{Kruijssen2009}. 

\subsection{A pure red HB cluster: NGC~104}
\begin{figure*}
\centering 
\includegraphics[scale=1,trim=8cm 13.5cm 9cm 6cm] {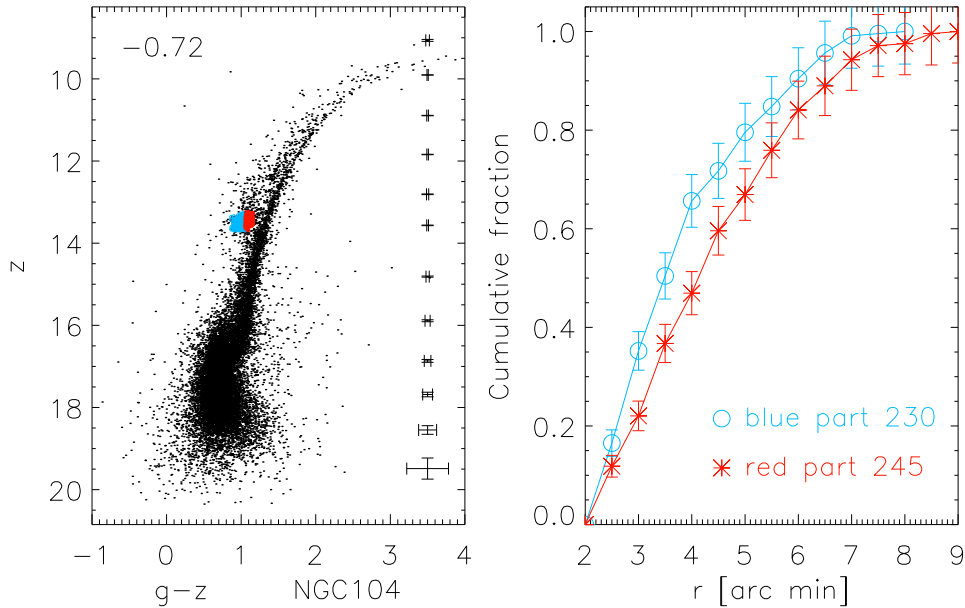} 
\caption{CMD of NGC104, a cluster lacking BHB and EHB, but with known multiple populations.  }
\label{fig:NGC104}
\end{figure*}

\cite{Gratton2013} demonstrated that only the reddest HB stars in NGC~104 (47 Tuc) can be considered as FG stars, the bluest ones are 
enriched in Na and depleted in O. In Fig.~\ref{fig:NGC104} we show the CMD and radial cumulative fractions for NGC~104. Although our 
"standard" approach is tailored to clusters with extended HBs, it is possible to make an arbitrary cut in 47 Tuc and study the cumulative radial 
distributions. We recover the \cite{Gratton2013} result and find that the blue part of the RHB is statistically more centrally concentrated than the 
red part (with $prob_{KS}=0.0001$). It further confirms the results of \cite{Nataf2011}, who found evidence for a centrally concentrated, He-rich 
SG. However it may be an ad hoc result and one should apply this method to metal-rich or red HB clusters with caution, although the success of 
this experiment suggests that our approach does select ranges where one or the other population is more significant.

\section{Summary} \label{sec:summary}Our data show that the distribution of stellar populations is more complex than
expected. While most GC formation scenarios predict that the second and further generations will have different
radial distribution from the primordial cluster populations, usually in the sense of being more centrally
concentrated, we find that there is little evidence that this is generally true, if we use HB stars as tracers
of He enriched population, a feature that should accompany the light element enhancement typical of
the secondary stellar generations. 

Although in some cases we expect that the stellar populations will have been thoroughly mixed by
dynamical evolution, including the possible effects of disk shocking, simulations of \cite{Vesperini2013} predict that original population gradients will not have been erased by the present time in many GCs.

In that respect, NGC~288, {NGC~362 and NGC~6218} are of specific interest. For these systems, 
the HB stars linked with the second generation are significantly less concentrated than the RHB stars. 
Photometric or spectroscopic follow-up studies are needed to confirm or disprove these radial distributions. 

One caveat on the present work is that we have to assume that the spread in colour on the HB in
each cluster is due to variations in helium abundance and can be related to the various stellar 
generations known to be present in these objects. While this seems to be a reasonable assumption
there are clusters where no helium variation is apparent on the HB, even though chemical anomalies
and multiple populations are present. 

Taken at face value, our results are not fully consistent with current enrichment scenarios dominated by AGB
stars or fast rotating massive stars, as these would generally produce more highly concentrated second
generation stars. Nevertheless, this points to the necessity of improving our theoretical understanding 
and modelling of multiple stellar populations in clusters, as well as defining a consensus tracer population.
Hubble Space Telescope photometry (e.g. \citealt{Piotto2015}) allows to extend our analysis into the 
cluster centres and could potentially provide additional clues. 

\section*{Acknowledgements}
We thankfully acknowledge the anonymous referee for constructive comments that improved the content of the present study. JV acknowledges the support of ESO through a studentship. JV and MB acknowledge the support of the Fund for Scientific Research Flanders (FWO-Vlaanderen). J.A.-G. acknowledges support from Proyecto Fondecyt Postdoctoral 3130552. The authors are grateful to CTIO for the hospitality and the dedicated assistance during the numerous observing runs.

\bibliographystyle{mnras}
\bibliography{references}

\begin{thebibliography}{63}
\expandafter\ifx\csname natexlab\endcsname\relax\def\natexlab#1{#1}\fi

\bibitem[{Arp} et~al.(1952){Arp}, {Baum} \& {Sandage}]{Arp1952}
{Arp} H.~C., {Baum} W.~A., {Sandage} A.~R., 1952, \aj, 57, 4

\bibitem[{Bastian} et~al.(2013){Bastian}, {Lamers}, {de Mink}, {Longmore},
  {Goodwin} \& {Gieles}]{Bastian2013}
{Bastian} N., {Lamers} H.~J.~G.~L.~M., {de Mink} S.~E., {Longmore} S.~N.,
  {Goodwin} S.~P., {Gieles} M., 2013, \mnras, 436, 2398

\bibitem[{Cardelli} et~al.(1989){Cardelli}, {Clayton} \&
  {Mathis}]{Cardelli1989}
{Cardelli} J.~A., {Clayton} G.~C., {Mathis} J.~S., 1989, \apj, 345, 245

\bibitem[{Carretta} et~al.(2010{\natexlab{a}}){Carretta}, {Bragaglia},
  {D'Orazi}, {Lucatello} \& {Gratton}]{Carretta2010d}
{Carretta} E., {Bragaglia} A., {D'Orazi} V., {Lucatello} S., {Gratton} R.~G.,
  2010{\natexlab{a}}, \aap, 519, A71

\bibitem[{Carretta} et~al.(2009){Carretta}, {Bragaglia}, {Gratton}
  et~al.]{Carretta2009}
{Carretta} E., {Bragaglia} A., {Gratton} R.~G., et~al., 2009, \aap, 505, 117

\bibitem[{Carretta} et~al.(2010{\natexlab{b}}){Carretta}, {Bragaglia},
  {Gratton} et~al.]{Carretta2010b}
{Carretta} E., {Bragaglia} A., {Gratton} R.~G., et~al., 2010{\natexlab{b}},
  \aap, 516, A55

\bibitem[{Carretta} et~al.(2006){Carretta}, {Bragaglia}, {Gratton}, {Leone},
  {Recio-Blanco} \& {Lucatello}]{Carretta2006}
{Carretta} E., {Bragaglia} A., {Gratton} R.~G., {Leone} F., {Recio-Blanco} A.,
  {Lucatello} S., 2006, \aap, 450, 523

\bibitem[{Carretta} et~al.(2011){Carretta}, {Lucatello}, {Gratton}, {Bragaglia}
  \& {D'Orazi}]{Carretta2011}
{Carretta} E., {Lucatello} S., {Gratton} R.~G., {Bragaglia} A., {D'Orazi} V.,
  2011, \aap, 533, A69

\bibitem[{Carretta} et~al.(2007){Carretta}, {Recio-Blanco}, {Gratton}, {Piotto}
  \& {Bragaglia}]{Carretta2007}
{Carretta} E., {Recio-Blanco} A., {Gratton} R.~G., {Piotto} G., {Bragaglia} A.,
  2007, \apjl, 671, L125

\bibitem[{Clayton}(1968)]{Clayton1968}
{Clayton} D.~D., 1968, {Principles of stellar evolution and nucleosynthesis}

\bibitem[{Dalessandro} et~al.(2014){Dalessandro}, {Massari}, {Bellazzini}
  et~al.]{Dalessandro2014}
{Dalessandro} E., {Massari} D., {Bellazzini} M., et~al., 2014, \apjl, 791, L4

\bibitem[{Dalessandro} et~al.(2011){Dalessandro}, {Salaris}, {Ferraro}
  et~al.]{Dalessandro2011}
{Dalessandro} E., {Salaris} M., {Ferraro} F.~R., et~al., 2011, \mnras, 410, 694

\bibitem[{D'Antona} et~al.(2005){D'Antona}, {Bellazzini}, {Caloi}, {Pecci},
  {Galleti} \& {Rood}]{Dantona2005}
{D'Antona} F., {Bellazzini} M., {Caloi} V., {Pecci} F.~F., {Galleti} S., {Rood}
  R.~T., 2005, \apj, 631, 868

\bibitem[{D'Antona} \& {Ventura}(2007)]{Dantona2007}
{D'Antona} F., {Ventura} P., 2007, \mnras, 379, 1431

\bibitem[{de Mink} et~al.(2009){de Mink}, {Pols}, {Langer} \&
  {Izzard}]{DeMink2009}
{de Mink} S.~E., {Pols} O.~R., {Langer} N., {Izzard} R.~G., 2009, \aap, 507, L1

\bibitem[{Decressin} et~al.(2010){Decressin}, {Baumgardt}, {Charbonnel} \&
  {Kroupa}]{Decressin2010}
{Decressin} T., {Baumgardt} H., {Charbonnel} C., {Kroupa} P., 2010, \aap, 516,
  A73

\bibitem[{Decressin} et~al.(2008){Decressin}, {Baumgardt} \&
  {Kroupa}]{Decressin2008}
{Decressin} T., {Baumgardt} H., {Kroupa} P., 2008, \aap, 492, 101

\bibitem[{Decressin} et~al.(2007){Decressin}, {Meynet}, {Charbonnel},
  {Prantzos} \& {Ekstr{\"o}m}]{Decressin2007}
{Decressin} T., {Meynet} G., {Charbonnel} C., {Prantzos} N., {Ekstr{\"o}m} S.,
  2007, \aap, 464, 1029

\bibitem[{D'Ercole} et~al.(2008){D'Ercole}, {Vesperini}, {D'Antona}, {McMillan}
  \& {Recchi}]{Dercole2008}
{D'Ercole} A., {Vesperini} E., {D'Antona} F., {McMillan} S.~L.~W., {Recchi} S.,
  2008, \mnras, 391, 825

\bibitem[{Dotter} et~al.(2010){Dotter}, {Sarajedini}, {Anderson}
  et~al.]{Dotter2010}
{Dotter} A., {Sarajedini} A., {Anderson} J., et~al., 2010, \apj, 708, 698

\bibitem[{Gratton} et~al.(2012){Gratton}, {Carretta} \&
  {Bragaglia}]{Gratton2012}
{Gratton} R.~G., {Carretta} E., {Bragaglia} A., 2012, \aapr, 20, 50

\bibitem[{Gratton} et~al.(2013){Gratton}, {Lucatello}, {Sollima}
  et~al.]{Gratton2013}
{Gratton} R.~G., {Lucatello} S., {Sollima} A., et~al., 2013, \aap, 549, A41

\bibitem[{Gratton} et~al.(2014){Gratton}, {Lucatello}, {Sollima}
  et~al.]{Gratton2014a}
{Gratton} R.~G., {Lucatello} S., {Sollima} A., et~al., 2014, \aap, 563, A13

\bibitem[{Gratton} et~al.(2015){Gratton}, {Lucatello}, {Sollima}
  et~al.]{Gratton2015}
{Gratton} R.~G., {Lucatello} S., {Sollima} A., et~al., 2015, \aap, 573, A92

\bibitem[{Harris}(1996)]{Harris1996}
{Harris} W.~E., 1996, \aj, 112, 1487

\bibitem[{Iannicola} et~al.(2009){Iannicola}, {Monelli}, {Bono}
  et~al.]{Iannicola2009}
{Iannicola} G., {Monelli} M., {Bono} G., et~al., 2009, \apjl, 696, L120

\bibitem[{Izzard} et~al.(2006){Izzard}, {Dray}, {Karakas}, {Lugaro} \&
  {Tout}]{Izzard2006}
{Izzard} R.~G., {Dray} L.~M., {Karakas} A.~I., {Lugaro} M., {Tout} C.~A., 2006,
  \aap, 460, 565

\bibitem[{Johnson} \& {Pilachowski}(2012)]{Johnson2012}
{Johnson} C.~I., {Pilachowski} C.~A., 2012, \apjl, 754, L38

\bibitem[{Joo} \& {Lee}(2013)]{Joo2013}
{Joo} S.-J., {Lee} Y.-W., 2013, \apj, 762, 36

\bibitem[{Krause} et~al.(2013){Krause}, {Charbonnel}, {Decressin}, {Meynet} \&
  {Prantzos}]{Krause2013}
{Krause} M., {Charbonnel} C., {Decressin} T., {Meynet} G., {Prantzos} N., 2013,
  \aap, 552, A121

\bibitem[{Krause} et~al.(2012){Krause}, {Charbonnel}, {Decressin}, {Meynet},
  {Prantzos} \& {Diehl}]{Krause2012}
{Krause} M., {Charbonnel} C., {Decressin} T., {Meynet} G., {Prantzos} N.,
  {Diehl} R., 2012, \aap, 546, L5

\bibitem[{Kravtsov} et~al.(2011){Kravtsov}, {Alca{\'{\i}}no}, {Marconi} \&
  {Alvarado}]{Kravtsov2011}
{Kravtsov} V., {Alca{\'{\i}}no} G., {Marconi} G., {Alvarado} F., 2011, \aap,
  527, L9

\bibitem[{Kruijssen} \& {Mieske}(2009)]{Kruijssen2009}
{Kruijssen} J.~M.~D., {Mieske} S., 2009, \aap, 500, 785

\bibitem[{Kunder} et~al.(2013){Kunder}, {Stetson}, {Cassisi}
  et~al.]{Kunder2013a}
{Kunder} A., {Stetson} P.~B., {Cassisi} S., et~al., 2013, \aj, 146, 119

\bibitem[{Lardo} et~al.(2011){Lardo}, {Bellazzini}, {Pancino}, {Carretta},
  {Bragaglia} \& {Dalessandro}]{Lardo2011}
{Lardo} C., {Bellazzini} M., {Pancino} E., {Carretta} E., {Bragaglia} A.,
  {Dalessandro} E., 2011, \aap, 525, A114

\bibitem[{Lee} et~al.(2007){Lee}, {Gim} \& {Casetti-Dinescu}]{Lee2007}
{Lee} Y.-W., {Gim} H.~B., {Casetti-Dinescu} D.~I., 2007, \apjl, 661, L49

\bibitem[{Mackey} \& {van den Bergh}(2005)]{Mackey2005}
{Mackey} A.~D., {van den Bergh} S., 2005, \mnras, 360, 631

\bibitem[{Marino} et~al.(2011){Marino}, {Villanova}, {Milone}
  et~al.]{Marino2011}
{Marino} A.~F., {Villanova} S., {Milone} A.~P., et~al., 2011, \apjl, 730, L16

\bibitem[{Milone}(2015)]{Milone2015}
{Milone} A.~P., 2015, \mnras, 446, 1672

\bibitem[{Milone} et~al.(2008){Milone}, {Bedin}, {Piotto} et~al.]{Milone2008}
{Milone} A.~P., {Bedin} L.~R., {Piotto} G., et~al., 2008, \apj, 673, 241

\bibitem[{Milone} et~al.(2014){Milone}, {Marino}, {Dotter} et~al.]{Milone2014a}
{Milone} A.~P., {Marino} A.~F., {Dotter} A., et~al., 2014, \apj, 785, 21

\bibitem[{Milone} et~al.(2009){Milone}, {Stetson}, {Piotto} et~al.]{Milone2009}
{Milone} A.~P., {Stetson} P.~B., {Piotto} G., et~al., 2009, \aap, 503, 755

\bibitem[{Monelli} et~al.(2013){Monelli}, {Milone}, {Stetson}
  et~al.]{Monelli2013}
{Monelli} M., {Milone} A.~P., {Stetson} P.~B., et~al., 2013, \mnras, 431, 2126

\bibitem[{Moni Bidin} et~al.(2012){Moni Bidin}, {Villanova}, {Piotto},
  {Moehler}, {Cassisi} \& {Momany}]{MoniBidin2012}
{Moni Bidin} C., {Villanova} S., {Piotto} G., {Moehler} S., {Cassisi} S.,
  {Momany} Y., 2012, \aap, 547, A109

\bibitem[{Mucciarelli} et~al.(2014){Mucciarelli}, {Lovisi}, {Lanzoni} \&
  {Ferraro}]{Mucciarelli2014}
{Mucciarelli} A., {Lovisi} L., {Lanzoni} B., {Ferraro} F.~R., 2014, \apj, 786,
  14

\bibitem[{Nataf} et~al.(2011){Nataf}, {Gould}, {Pinsonneault} \&
  {Stetson}]{Nataf2011}
{Nataf} D.~M., {Gould} A., {Pinsonneault} M.~H., {Stetson} P.~B., 2011, \apj,
  736, 94

\bibitem[{Piotto} et~al.(2007){Piotto}, {Bedin}, {Anderson} et~al.]{Piotto2007}
{Piotto} G., {Bedin} L.~R., {Anderson} J., et~al., 2007, \apjl, 661, L53

\bibitem[{Piotto} et~al.(2015){Piotto}, {Milone}, {Bedin} et~al.]{Piotto2015}
{Piotto} G., {Milone} A.~P., {Bedin} L.~R., et~al., 2015, \aj, 149, 91

\bibitem[{Renzini}(1998)]{Renzini1998}
{Renzini} A., 1998, \aj, 115, 2459

\bibitem[{Salaris} et~al.(2008){Salaris}, {Cassisi} \&
  {Pietrinferni}]{Salaris2008}
{Salaris} M., {Cassisi} S., {Pietrinferni} A., 2008, \apjl, 678, L25

\bibitem[{Salaris} et~al.(2013){Salaris}, {de Boer}, {Tolstoy}, {Fiorentino} \&
  {Cassisi}]{Salaris2013}
{Salaris} M., {de Boer} T., {Tolstoy} E., {Fiorentino} G., {Cassisi} S., 2013,
  \aap, 559, A57

\bibitem[{Sandage}(1953)]{Sandage1953}
{Sandage} A.~R., 1953, \aj, 58, 61

\bibitem[{Saviane} et~al.(1998){Saviane}, {Piotto}, {Fagotto}, {Zaggia},
  {Capaccioli} \& {Aparicio}]{Saviane1998}
{Saviane} I., {Piotto} G., {Fagotto} F., {Zaggia} S., {Capaccioli} M.,
  {Aparicio} A., 1998, \aap, 333, 479

\bibitem[{Sbordone} et~al.(2011){Sbordone}, {Salaris}, {Weiss} \&
  {Cassisi}]{Sbordone2011}
{Sbordone} L., {Salaris} M., {Weiss} A., {Cassisi} S., 2011, \aap, 534, A9

\bibitem[{Sollima} et~al.(2007){Sollima}, {Ferraro}, {Bellazzini}, {Origlia},
  {Straniero} \& {Pancino}]{Sollima2007}
{Sollima} A., {Ferraro} F.~R., {Bellazzini} M., {Origlia} L., {Straniero} O.,
  {Pancino} E., 2007, \apj, 654, 915

\bibitem[{Stetson}(1987)]{Stetson1987}
{Stetson} P.~B., 1987, \pasp, 99, 191

\bibitem[{Vanderbeke} et~al.(2014{\natexlab{a}}){Vanderbeke}, {West}, {De
  Propris} et~al.]{Vanderbeke2014a}
{Vanderbeke} J., {West} M.~J., {De Propris} R., et~al., 2014{\natexlab{a}},
  \mnras, 437, 1725

\bibitem[{Vanderbeke} et~al.(2014{\natexlab{b}}){Vanderbeke}, {West}, {De
  Propris} et~al.]{Vanderbeke2014b}
{Vanderbeke} J., {West} M.~J., {De Propris} R., et~al., 2014{\natexlab{b}},
  \mnras, 437, 1734

\bibitem[{Vesperini} et~al.(2013){Vesperini}, {McMillan}, {D'Antona} \&
  {D'Ercole}]{Vesperini2013}
{Vesperini} E., {McMillan} S.~L.~W., {D'Antona} F., {D'Ercole} A., 2013,
  \mnras, 429, 1913

\bibitem[{Villanova} et~al.(2014){Villanova}, {Geisler}, {Gratton} \&
  {Cassisi}]{Villanova2014}
{Villanova} S., {Geisler} D., {Gratton} R.~G., {Cassisi} S., 2014, \apj, 791,
  107

\bibitem[{Villanova} et~al.(2012){Villanova}, {Geisler}, {Piotto} \&
  {Gratton}]{Villanova2012}
{Villanova} S., {Geisler} D., {Piotto} G., {Gratton} R.~G., 2012, \apj, 748, 62

\bibitem[{Villanova} et~al.(2009){Villanova}, {Piotto} \&
  {Gratton}]{Villanova2009}
{Villanova} S., {Piotto} G., {Gratton} R.~G., 2009, \aap, 499, 755

\bibitem[{Walker}(1998)]{Walker1998}
{Walker} A.~R., 1998, \aj, 116, 220

\end{thebibliography}

\label{lastpage}
\end{document}